# ARTICLE

# Unveiling nutrient flow mediated stress in the plant roots using on-chip phytofluidic device

Kaushal Agarwal,*[a] Sumit Kumar Mehta [b] and Pranab Kumar Mondal [a,b]

The initial emergence of the primary root from a germinating seed is a pivotal phase that influences a plant's survival. Abiotic factors such as pH, nutrient availability, and soil composition significantly affect root morphology and architecture. Of particular interest is the impact of nutrient flow on thigmomorphogenesis, a response to mechanical stimulation in early root growth, which remains largely unexplored. This study explores the intricate factors influencing early root system development, with a focus on the cooperative correlation between nutrient uptake and its flow dynamics. Using physiologically relevant, portable, and cost-effective microfluidic system for the controlled fluid environments offering hydraulic conductivity comparable to that of the soil, this study analyzes the interplay between nutrient flow and root growth post-germination. Emphasizing the relationship between root growth and nitrogen uptake, the findings reveal that nutrient flow significantly influences early root morphology, leading to increased length and improved nutrient uptake, varying with the flow rate. The experimental findings are supported by stress-related fluid flow-root interaction simulations and quantitative determination of nitrogen uptake using the Total Kjeldahl Nitrogen (TKN) method. The microfluidic approach offers novel insights into plant root dynamics under controlled flow conditions, filling a critical research gap. By providing a high-resolution platform, this study contributes to the understanding of how fluid-flow assisted nutrient uptake and pressure affect root-cell behavior, which, in turn, induces mechanical stress leading to thigmomorphogenesis. The findings hold implications for comprehending root responses to changing environmental conditions, paving the way for innovative agricultural and environmental management strategies.

## Introduction

The primary root that emerges first from the germinating seed functions as the physical anchor of the plant [1]. It is responsible for absorbing water and nutrients from its microhabitat and facilitating the transport of these resources to other parts of the plant [2]. The emergent primary root encounters challenges as it must grow and penetrates the heterogenous soil substrate. Thus, the early stages of root growth are a critical window for the plant's survival [1]. Early growth and development of the root system are influenced by various abiotic factors such as nutrient supply, pH, total cationic concentration, aeration and soil temperature, humidity, strength, composition, and water content [3–6]. Besides these, external hydrostatic and hydrodynamic pressure are also known to have a non-trivial impact on root architecture [7,8]. Nutrient supply can potently affect root growth, morphology, and distribution in the substrate [9]. The flow of nutrient supply or water provides plant roots with a suitable level of mechanical stimulation that subsequently promotes root growth. The root growth as well as its branching are adjusted to optimize the nutrient provision to the plant. Among the nutrients required by plants, the availability of the macronutrient, nitrogen, plays a crucial role in plant growth and development. Nitrogen is known to play a vital role in the modulation of root architecture, root biomass, and growth. It may be mentioned here that a localized nutrient supply can stimulate root growth in an optimized fashion [10–12]. However, the experimental setup for studies involving roots typically utilizes macroscopic vessels and containers, which necessitates the development of methods for large-scale phenotyping [13,14]. This, in turn, results in challenges when studying the dynamics of plant root systems.

The advent of microfluidics, that refers to the study of flow dynamics, characterization of fluid properties, and its control within micrometer-sized structures [13], has enabled to recreate *in vitro* environments for cell studies [14]. In the realm of plant studies, however, only a limited number of microdevices have been developed so far among which the majority of literature deals with root-bacteria interactions [15,16], hormonal signaling [17], and the growth of pollen tubes [2,18–21]. Nevertheless, as witnessed from the referred literature [2,14,17,22], the exploration of *in vitro* investigations into plant root dynamics and its real-time analysis using on-chip microfluidic platforms, has remained unexplored until recently.

[a.] School of Agro and Rural Technology, Indian Institute of Technology Guwahati, Guwahati- 781039, India

[b.] Microfluidics and Microscale Transport Processes Laboratory, Department of Mechanical Engineering, Indian Institute of Technology Guwahati, Guwahati- 781039, India

*Corresponding author; E-mail: pranabm@iitg.ac.in, mail2pranab@gmail.com (P. K. Mondal) Phone: +91-361-2583435 (Office); Fax: +91-361-2582699

While the influence of abiotic factors such as pH and nutrient availability on root morphology is well-documented [23–32], the novel aspect of our study is the examination of the specific effects of mechanical stimuli resulting from nutrient flow on thigmomorphogenesis by using on-chip microfluidic device, which has not been previously explored. Thigmomorphogenesis is a very real phenomenon in plants which affects many aspects of plant growth and development [33]. It is important to note that the nature or extent of the response depends on the species or variety, as well as the physiological stage of the plant when it is stimulated. No literature has been found yet which investigates the influence of nutrient flow and its dynamics through microfluidics in plant roots. Hence, there is a need for perfusion devices that can be used to study the root dynamics and the interplay of nutrient solution at a high resolution, eliminating the challenges associated with specimen handling.

Taking into account the constraints related to current plant root studies, herein, we describe and present the design of a physiologically relevant, portable, easy-to-use, and low-cost microfluidic setup. The fluidic configuration considered in this endeavor offers hydraulic conductivity of equal order to that of the soil in real scenarios [34,35]. In this study, we employ *Brassica juncea* (Pusa Jaikisan), a fast growing, non-competitive, oilseed, dicot crop, having an effective root diameter within micrometer range [36–38]. The goal is to explore how nutrient flow conditions impact the growth of the developing root during post-germination stages. Also, we demonstrate the uptake of one of the vital macronutrients, nitrogen, and its correlation with plant root growth. This work provides a comprehensive description of the experimental setup and the reasoning behind its design choices. The presented experimental findings are validated through three key proofs of concept: (I) the estimation of generated stress through the numerical simulations of nutrient flow-root interaction within the microfluidic channel using mathematical framework, (II) determination of nitrogen uptake using the Total Kjeldahl Nitrogen method [39] and, (III) the influence of fluid-flow-assisted nutrient uptake and fluid pressure on root-cells using free-hand anatomical sections. here.

## Materials and methods

### Plant material: Seed sources and surface sterilization

Mature seeds of *Brassica juncea* (Pusa Jaikisan) were collected from ICAR-IARI, Regional Station, Karnal, Haryana, India. The seeds were stored in air-tight falcon tubes at room temperature. To remove dirt and microbial contaminants present on the seed surface, the seeds were surface sterilized sequentially with 70% ethyl alcohol (v/v) for 30 seconds and 4% Sodium hypochlorite solution (w/v) for 5 minutes. They were then thoroughly rinsed with autoclaved double-distilled water for 1 minute (5 times). To surface-dry, the seeds were placed on UV-sterilized blotting paper in a petri plate. The sterilization process was carried out inside the laminar-airflow cabinet, after which the seeds were used for *in vitro* germination.

### Germination setup and conditions

Following the surface drying process, the sterilized seeds were finally placed within a sterile 90 mm petri plate (Tarsons™) containing filter paper (Grade 1, Whatman®), wetted with sterile distilled water (cf. Fig. 1a-i). Seeds were arranged in a 5×5 pattern, forming columns and rows across the plate and sealed with laboratory film (Parafilm M®). Subsequently, the petri plate containing the seeds was placed in a germination chamber maintained at a temperature of 26 °C and a relative humidity of 75% under 16/8 hours (light/dark) photoperiod with a light intensity of 3000 lux provided by white fluorescent tubes [13,14,17,40–44].

### Fabrication of PRFD and PRFS Assembly

For observing and analyzing the real-time root growth dynamics of *Brassica juncea*, we fabricated a Plant Root Fluidic Device (PRFD), as shown in the inset of Fig. 1b. Each PRFD was fabricated with dimensions: 50 mm (length) × 5 mm (breadth) × 30 mm (height), with a channel of effective length 45 mm. For the study, eight such PRFDs, collectively known as PRFS, were used in experiments. The master mold of the device was designed using AutoCAD® software (Autodesk™, USA) and then 3D printed with ABS (Acrylonitrile butadiene styrene) thermoplastics. The device was fabricated using a vinyl-terminated polydimethylsiloxane (SYLGARD™ 184 Silicone Elastomer, Sigma Aldrich®, USA) substrate, following an inexpensive wire-drawing method with adaptations from published methods [45–47]. It was then soft-baked at 45 °C for 10 hours in a hot air oven (IKON™ Instruments, India). The silicon polymer formed a negative mold of the device, which was carefully separated from the master mold to create the final device of interest. Locally purchased brass wire with a cross-section of 0.8 mm × 0.8 mm was used to mimic the horizontal channel for root growth in the PRFD. Pipette tips with definite length and diameter were used to hold the seedling vertically and create inlet and outlet connections for fluid exchange. Here, it should be mentioned that the PDMS-based channels are biocompatible and have no influence on the growth of the plant.

### Seedling Integration in the PRFD

At 2 DAG (Day After Germination), under sterile conditions, the seedlings were taken from the petri plate and placed inside the vertically positioned pipette tip connected with the horizontal channel (also, known as plant inlet in Fig. 1a-ii), which was 15 mm away from the inlet port of the PRFD. The procedural steps of seedling transfer from the petri plate to each PRFD are schematically demonstrated in Fig. 1a-ii and iii. This placement

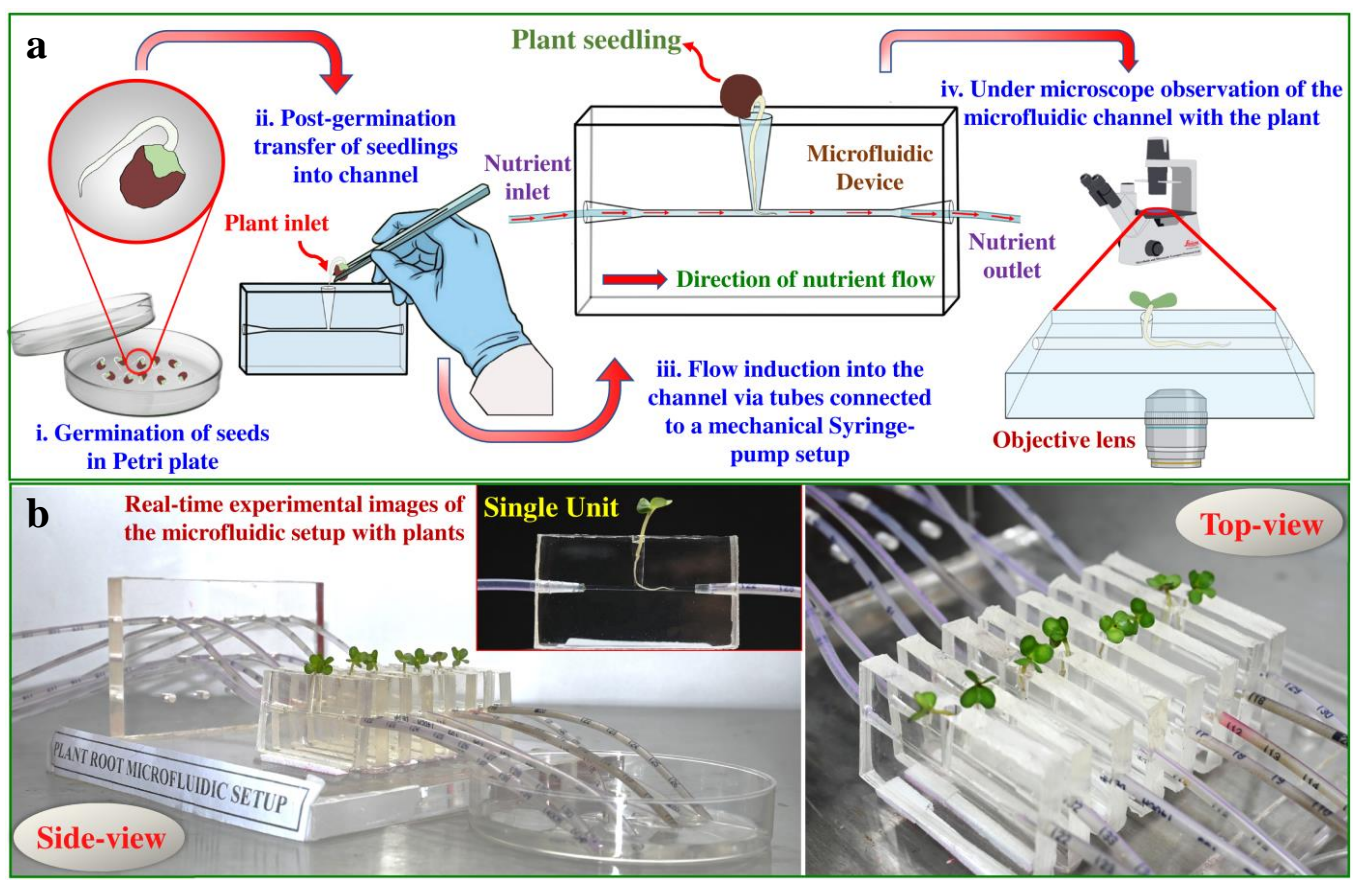

**Fig. 1** (a) Schematic representation of the experimental methods in the microfluidic study conducted. Seeds were germinated under sterile conditions and then carefully transferred to the microfluidic setup. Nutrient flow within the channel was introduced using a mechanical syringe pump with a definite flow rate ($Q$). Syringes containing the nutrient were connected to the channels using sterile baby-feeding tubes, serving as the inlet for the channels. Upon successful integration of the root system with the microfluidic system, the root was observed under the microscope for 12 hours; micrographs were taken at 2-hour intervals without disturbing the plant system. (b) Experimental images of the setup showing the nutrient being pumped into the microfluidic setup in side-view (left), single microfluidic channel as an inset of the figure; side-upper, top-view of the plant-integrated microfluidic setup (right). Note: here, images are not to scale allowed the growing roots to conveniently enter the horizontal channel underneath. Each channel in the PRFD, configured with a plant, was supplied with full strength MS media for 12 hours via the inlet port, ensuring definite and distinct fluid-flow conditions while maintaining consistent environmental conditions for all.

**Real-Time Micrography of the Root Dynamics**
For real-time micrography of the root growing inside each PRFD, the device was placed under the Leica® DMI3000 M inverted microscope coupled with a high-speed camera (Phantom, Vision Research Inc. USA; Model: Miro LAB320) boasting a resolution of 1900 ×1200 pixels, as represented schematically in Fig. 1a-iv. The objectives used were HC FL PLAN 2.5×/0.07, 5 ×/0.12 N PLAN EPI, and 10×/0.25. N PLAN (Leica). Micrographs for each case were imaged every 2 hours using the aforementioned objectives over a 12-hour period.

**Image Acquisition and Data Analysis**
Image analysis was performed using ImageJ/Fiji (Open Source) Software to measure the lengths of the roots under different flow rate conditions. Subsequently, data analysis was carried out using MATLAB® R2022a (The MathWorks Inc.).

**AFM Assisted Imaging and Measurement of Young's Modulus**
To determine the Young's Modulus of the growing root (soft tissue), the procedure followed in this study is briefly outlined here for the sake of completeness. The root sample was placed on thoroughly cleaned glass slides using sterilized forceps and air-dried overnight. Subsequently, the air-dried root sample was imaged using an Asylum Research MFP-3D-BIO (Asylum Research of Oxford Instruments®, UK) Atomic Force Microscope (AFM). The slide was fixed temporarily onto a metal disc and then placed in the AFM for scanning. AFM imaging was conducted in Nanoindentation mode using an AC160TS-R3 probe (Oxford Instruments®, UK). The Young's modulus was determined by fitting curves using the Hertz Model as depicted in Fig. 4a, while the sample preparation steps, from fixation to imaging, together with the sample used for this part are shown in the inset of Fig. 4a. Likewise, the point-wise modulus and indentation curves are shown in Figs. 4B and its inset, respectively. The analysis of AFM images and the calculation for Young's modulus were performed using both AtomicJ (Open Source) and NanoScope Analysis (Bruker, Billerica, MA, USA) software.

**Anatomical Analysis of the Root Sections: Staining, Imaging, and Analysis**

After 12 hours of observation in the microfluidic channel, the plant was carefully taken out and free-hand sections of the delicate root were done by embedding the root in agar medium. The thin sections were then stained with safranin, followed by a series of washes with graded alcohol and water. The thin sections (stained samples) were observed under a microscope, and images of the sections were analysed using ImageJ/Fiji Software (Open Source).

**Estimation of Nitrogen Uptake**

Whole plant samples from each experimental case were collected and oven-dried for 24 hours. Subsequently, 200 mg of each sample was used for the estimation of nitrogen by the TKN (Total Kjeldahl Nitrogen) method [39] using TKN Analyzer (KES 08L R, Pelican Equipment, India).

**Statistical analysis of Root Growth Dynamics**

The data in this study have been expressed as mean values ± standard deviation (S.D.). Data were analyzed with IBM® SPSS® version 29.0 (Armonk, NY: IBM Corp) for significance using one-way analysis of variance (ANOVA), followed by the Duncan's multiple range test (DMRT) for contrasting differences. A significance level of $p_r$ < 0.05 was considered significant, unless otherwise specified, where $p_r$ defines the probability and determines the likelihood that any observed variation between groups is a result of random occurrences.

## Results and Discussion

**Plant Root Fluidic System (PRFS) and Experimental Setup**

The seedlings of *Brassica juncea*, after germination (Fig. 1a-i), were placed in the Plant Root Fluidic Device (PRFD) as depicted schematically in Fig. 1a-ii & iii. The actual experimental setup consisted of eight PRFDs, each having a designated microfluidic channel with effective dimensions of 45 mm (length) × 0.8 mm (breadth) × 0.8 mm (height), collectively termed as Plant Root Fluidic System (PRFS). Each vertically standing PRFD (precisely, each microfluidic channel) in the PRFS, shown in Fig. 1b as top- and side-view, was connected to individual 10 mL syringe (NIPRO, India) containing MS media (Murashige & Skoog Medium, PT021, HIMEDIA®, India) through individual inlet ports using sterile feeding tubes with an effective length and diameter of 500 mm and 2.70 mm, respectively. The syringes were, in turn, connected to a syringe pump (New Era Pump, India) to supply an effective and continuous flow of the nutrient medium/solution through the channel of each PRFD. Henceforth, for the sake of convenience, we will simply write channel instead of PRFD. The outlet ports of each channel were extended to the collection sump using feeding tubes and straight connectors as seen in the side-view of Fig. 1b. To avoid contamination during the plant growth period, the entire experimental setup was placed in a UV-sterilized, closed plant growth chamber maintained at a temperature of 26 °C and 75% relative humidity, with a 16-hour light/8-hour dark cycle.

**Fluid-Flow Configuration**

Following the integration of the seedling into the PRFDs, as described in the '*Materials and Method*' section, each seedling was allowed to grow in a no-flow condition until its root entered the horizontal channel from the vertically placed pipette tip. Once the root entered the channel, syringe pumps were used to infuse nutrient media to nourish the growing roots. Fluid-flow inside the channel was set at various rates: 0.05 mL/hr, 0.1 mL/hr, 0.2 mL/hr, 0.4 mL/hr, 0.6 mL/hr, 0.8 mL/hr, 1.0 mL/hr and 1.2 mL/hr. The chosen window of flow rates complies with the physically permissible range typically considered in this paradigm [48–50]. For each flow rate, the flow direction was maintained in the positive direction of the root growth (towards channel outlet). An additional no-flow condition was maintained as a control for each flow rate configuration. The setup remained inside the sterile growth chamber and was observed for 12 hours after the root entered the horizontal channel. Each experiment was replicated three times on three separate instances.

**Hydraulic Conductivity of PRFD and Soil**

We attempted to estimate the hydraulic conductivity of the flow inside the developed PRFD and compared it with the hydraulic conductivity of soil available in the referred literature [34,35]. We undertook this endeavour to ascertain whether the flow rate considered through the PRFD during experiments in this study mimics practical scenarios. The instantaneous flux, or flow rate per unit cross-sectional area, can be expressed using Darcy's law: $q = (K\Delta p)/\mu L$. Here, $K$, $\Delta p$, $\mu$ and $L$ stand for the soil's permeability (measured in m²), pressure drop, dynamic viscosity, and length of PRFD, respectively. Expressing it as $q = (K\rho g \Delta h)/\mu L$ or $q = (K\rho g/\mu)(\Delta h/L)$ using $p = \rho g h$, where $h$ is the pressure head, the term $K\rho g/\mu$ is denoted by the letter $\kappa$ and stands for hydraulic conductivity (in m/s). It should be noted that the hydraulic conductivity of the PRFD is expected to be in the order of $O(\kappa) \sim 10^{-4}$ m/s based on the pressure drop and flow rate per unit cross-sectional area. For a flow rate of 0.05 mL/hr (equivalent to 1.38 × 10$^{-11}$ m³/s), with a rectangular cross-section of PRFD of 64 × 10$^{-8}$ m², the flux is calculated as $q$ = 3.70 × 10$^{-5}$ m/s. Using $\mu$ = 10$^{-3}$ Pa·s, $\rho$ = 1000 kg/m³, $L$ = 45 × 10$^{-3}$ m and the pressure drop estimated from numerical simulation, $\Delta p$ = 2.6387 × 10$^{-5}$ Pa, the hydraulic conductivity $\kappa$ (=$K\rho g/\mu$) is calculated as $3.625 \times 10^{-4}$ m/s, with $g$ = 9.8 m/s². Similarly, for a flow rate of 1.2 mL/hr, the value of $\kappa$ is obtained as $3.654 \times 10^{-4}$ m/s, with $g$ = 9.8 m/s$^2$. Notably, research in the referred literature indicates that the biochar-rich soil utilized in agriculture has hydraulic conductivity in the range of $O(\kappa) \sim 10^{-4}$ m/s [34,35]. This order analysis justifies that the chosen window of flow rate through the PRFD meets practical relevance in the established setup/configuration.

**Mathematical Model for Nutrient Flow-Root Interaction**

We here take an effort to estimate the pressure acting on the root surface due to flow loading and the corresponding stress being developed inside the root. To achieve this, we use the finite element framework of COMSOL Multiphysics™ [51] to solve for the flow field and deformation inside the root under the influence of flow loading. The steady-state transport equations governing the flow field ($\mathbf{u}$) and deformation of the root in the configuration consistent with the present experimental setup are as follows [52]:

$$\nabla \cdot (\mathbf{u}) = 0 \tag{1}$$

$$\rho(\mathbf{u} \cdot \nabla)\mathbf{u} = -\nabla p + \mu \nabla^2 \mathbf{u} \tag{2}$$

$$\nabla \bar{\sigma} = 0; \bar{\sigma} = 2\mu_L \bar{\varepsilon} + \lambda_L tr(\bar{\varepsilon}) I \tag{3}$$

The parameters $\rho$, $\mu$ and $p$ in Eq. (2) are, respectively, the density, viscosity, and pressure of the flowing fluid: in this case, nutrient solution. The first and second Lame parameters appearing in Eq. (3) are $\lambda_L = \nu E/(1+\nu)(2\nu-1)$ and $\mu_L = E/2(1+\nu)$, respectively [52]. The interfacial boundary condition is $\bar{\sigma} \cdot \mathbf{n}_{s,\Gamma} = -(\mu \nabla \mathbf{u}) \cdot \mathbf{n}_{f,\Gamma}$; where $\bar{\sigma}$ and $\bar{\varepsilon}$ are the stress and Green–Saint Venant strain tensor in the root; $\mathbf{n}$ is the normal unit vector to the interface, $\Gamma$; suffixes $s$ and $f$ denote solid root and nutrient fluid domains, respectively. The density and viscosity of the nutrient solution, which exhibits properties similar to water, are taken as $\rho$ = 1000 kg/m$^3$ and $\mu$ = 0.001 Pa·s, respectively. As evident from Eq. (3), the Young's modulus of the growing root is a critical parameter to estimate the deformation of the root under external forces (flow loading). Unfortunately, no prior literature reporting the mechanical properties, including the Young's modulus, of the same plant is available. To address this gap, we attempt to experimentally determine the Young's modulus of our sample plant root (*Brassica juncea*) using Atomic Force Microscopy (AFM), in accordance with the Hertz fit model, as described in the '*Materials and Method*' section. The estimated values of Young's modulus and Poisson's ratio for the growing root are obtained as 4.1625 MPa and 0.49, respectively. For the sake of conciseness in the presentation and to maintain focus of the present analysis, we do not explicitly discuss here the measurement method. However, interested readers may refer to the seminal studies available in this paradigm for this part [53–61]. For the simulations of the underlying flow field, we apply a no-slip boundary condition at the liquid-solid interface and assume a fully developed velocity field at the inlet to solve Eqs. (1) and (2). It is worth mentioning here that the effect of root deformation on fluid flow is neglected (i.e., one-way coupling) due to the weaker flow loading.

**PRFS Facilitates Real-Time Imaging of Root Micrograph**

Morphological analysis of the data obtained from micrography of the roots revealed that the variations in the flow rate of the nutrient media in the microfluidic channel affected root length, as shown schematically in Fig. 2a and as experimental micrographs inset in Fig. 2b. As evident from Regime-I of Fig. 2b, and Fig. 2c, the root length, in contrast to the no-flow scenario (control), increased significantly with the increase in the flow rate from 0.05 mL/hr (root length= 5.87 ± 0.373 mm) to 0.4 mL/hr (root length= 11.16 ± 0.378 mm). Interestingly, as observed in the analyzed results, any further increase in flow rate beyond 0.4 mL/hr (i.e., 0.6 mL/hr, 0.8 mL/hr, 1.0 mL/hr and 1.2 mL/hr.) resulted in a declining trend in root length (Regime-II of Fig. 2b, and Fig. 2d). Notably, the mean values marked with the same letter (s) did not differ significantly at $p_r \leq 0.05$ according to Duncan's multiple range test, where $p_r$ defines the probability and determines the likelihood that any observed variation between groups is a result of random occurrences (Fig. 2e). We refer to the flow rate of 0.4 mL/hr, at which the root length is maximum, as the "optimum flow rate". The root length data obtained at a flow rate of 0.6 mL/hr (root length= 9.29 ± 0.392 mm) shows a steep decrease compared to the root length at the optimum flow rate, i.e., 0.4 mL/hr. The intricate competition between hydrodynamic stress (mechanical stress) inevitably associated with the nutrient flow and corresponding nutrient uptake is deemed pertinent to reveal the basic morphogenesis of root length architecture with a change in flow rate. The reduced root length seems to result from increased stress on growing roots due to the hydrodynamic stimulation, as can be seen from our simulated data. However, it is important to note that in flow conditions, the observed root length consistently exceeded that in no-flow conditions. This can be primarily attributed to progressive nitrogen uptake by the growing root with nutrient flow rate, which indeed counteracts the stress experienced by them. We will discuss these aspects in greater detail in the later part of this article from the perspective of histology, biochemical analysis, and simulations.

**Anatomical Section of Root and Cellular Structure**

The analysis of micrographs from free-hand anatomical cross-section of roots under different flow rates (0.05 mL/hr, 0.4 mL/hr, and 1.2 mL/hr) reveals a peculiar change in the count and size of the cortical cells beneath the epidermal layer of the root sections. Specifically, the cortical cell count ($C_c$) increases with the enhancement in flow rate. For the roots under flow conditions consistent with the rate of 0.05 mL/hr ($Q_1$), 0.4 mL/hr ($Q_2$), and 1.2 mL/hr ($Q_3$), $C_c$ is found to be 68±3[a], 73±2[a] and 85±3[b], respectively (Fig. 3a). The mean values for $C_c(Q_1)$ and $C_c(Q_2)$, marked with the same letter, do not exhibit significant differences between themselves. Whereas $C_c(Q_3)$ shows a marked difference in $C_c$ compared to both $C_c(Q_1)$ and $C_c(Q_2)$, at a significance level of $p_r \leq 0.05$ as determined by Duncan's multiple range test (Fig. 3b). Notably, the higher flow rate ($Q_3$) exhibits a significant increase in $C_c$ compared to the lower flow rates ($Q_1$) and ($Q_3$). In contrast, $C_c$ for $Q_1$ and $Q_2$ shows a smaller variation in the number of cells, suggesting a change in $C_c$ with an increase in flow rate. Furthermore, we calculated the average area of cortical cells ($A_c$) for the flow rates $Q_1, Q_2$ and $Q_3$ using ImageJ as 0.1164±0.0286 mm², 0.1115±0.0296 mm² and 0.0912±0.0269 mm², respectively, a portion of which is shown

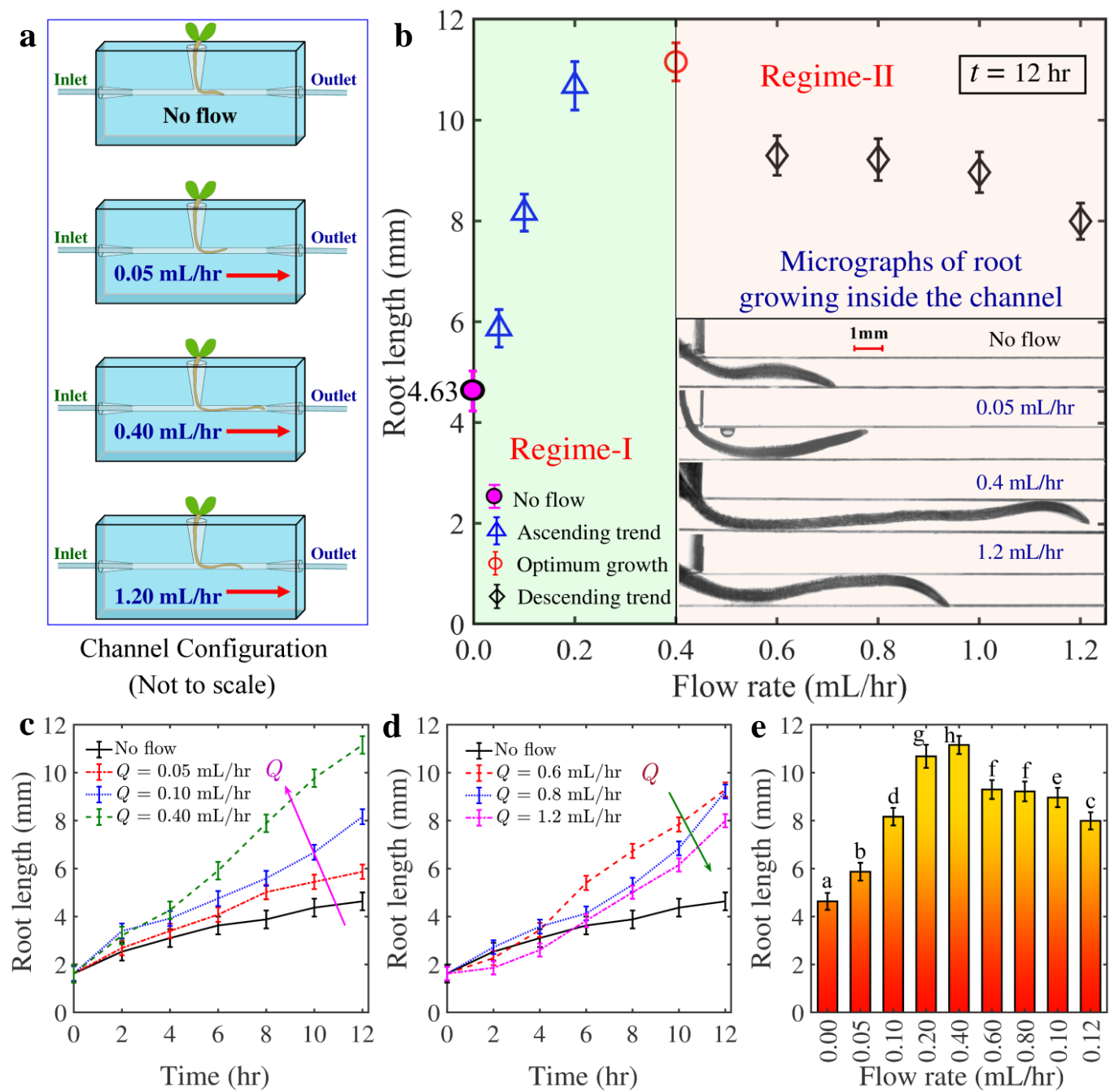

**Fig. 2** (a) Schematic representation of channel configuration for $Q$ = 0.05 mL/hr, $Q$ = 0.4 mL/hr and $Q$ = 1.2 mL/hr and no flow state. Red arrow (→) represents the direction of the flow of the nutrient through the channel. (b) Variation in the length of the root growing inside the microfluidic channel with respect to different flow rates ($Q$) at t = 12 hr. The two regimes, namely, Regime-I and Regime-II depict the change in root length in mm. The pink circle (⬤) denotes the length of the root (4.63 ± 0.34 mm) in the no-flow condition. The red marker (⬥) denotes the root length at optimum flow rate, 0.4 mL/hr. Regime-I denotes the increase in root length as $Q$ increases from 0.05 mL/hr to 0.4 mL/hr. Regime-II denotes the decrease in root when $Q$ is increased beyond 0.4 mL/hr up to 1.2 mL/hr. Micrographs in Regime-B show the root length variation for the no-flow condition, $Q$ = 0.05 mL/hr, $Q$ = 0.4 mL/hr and $Q$ = 1.2 mL/hr at t = 12 hr. Different markers are used to denote the ascending trend ($Q$ = 0.05 mL/hr to 0.2 mL/hr), descending trend ($Q$ = 0.6 mL/hr to 1.2 mL/hr) and optimum growth condition ($Q$ = 0.4 mL/hr). (c) Increasing variation in the length of the root growing inside the microfluidic channel for different $Q$ with respect to time. (d) Decreasing variation in the length of the root growing inside the microfluidic channel for different $Q$ with respect to time. In both variations, increasing as well as decreasing, the length of the root for different $Q$ is greater than that in the no-flow condition. (e) Variations in the average root length at different flow rates. The graphs were plotted using MATLAB® R2022a, and the statistical significance of the data was assessed through one-way ANOVA using IBM® SPSS® software. The mean values marked with the same letter (s) do not exhibit significant differences at a significance level of $p_r \leq 0.05$, as determined by Duncan's multiple range test, where $p_r$ defines the probability and determines the likelihood that any observed variation between groups is a result of random occurrences.

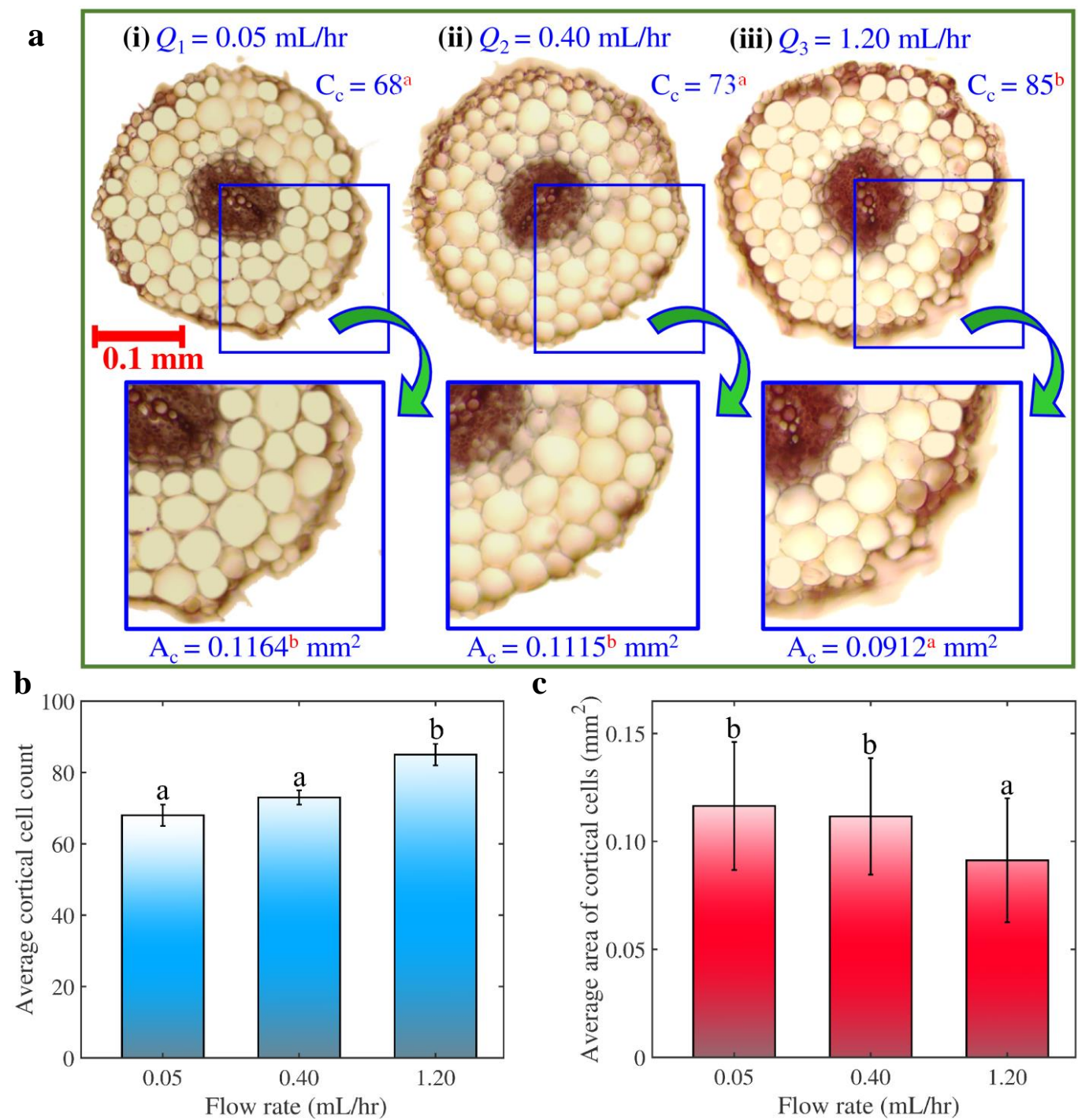

**Fig. 3** (a) Anatomical sections of the root depicting cell distribution and cortical cell count (Cc) for (i) $Q$ = 0.05 mL/hr, (ii) $Q$ = 0.4 mL/hr and (iii) $Q$ = 1.2 mL/hr, respectively. The zoomed view depicts detailed information about a portion of cortex beneath the epidermal layer, and the average area of cortical cells for each root section (Ac) is denoted below the zoomed view boxes. (b) Variation in the average cortical cell count (Cc) of the flow rates $Q$ = 0.05 mL/hr, $Q$ = 0.4 mL/hr and $Q$ = 1.2 mL/hr. (c) Variations in the average area (Ac) of cortical cells at different flow rates $Q$ = 0.05 mL/hr, $Q$ = 0.4 mL/hr and $Q$ = 1.2 mL/hr. The graphs were plotted using MATLAB® R2022a and the significance of the data was verified performing one-way ANOVA using IBM® SPSS® software. The mean values marked with the same letter (s) do not differ significantly at p_r≤0.05 according to Duncan's multiple range test.

as zoomed-in-view images in Fig. 3a. The mean values for $A_c(Q_1)$ and $A_c(Q_2)$ do not differ significantly between themselves. Whereas $A_c(Q_3)$ exhibits a difference in $A_c$ compared to both $A_c(Q_1)$ and $A_c(Q_2)$, at $\boldsymbol{p_r \leq 0.01}$ according to Duncan's multiple range test as seen in Fig. 3c. The data was used to deduce the reduction (R) percentage in the area of cells in the cortex for flow rates $Q_2$ and $Q_3$ relative to the lower flow rate $Q_1$ (as graphically shown in the inset of Fig. 6a in the upcoming section). The cortical cell distribution count and cell size reduction parameters are indicative of the adaptive nature of the roots to the stress experienced by the varying fluid flow rate inside the fluidic channel. We have elaborated on this aspect in the next section.

### Simulating Nutrient Flow-Root Interaction and Uptake

The effective growth of the plant root in the nutrient-saturated microenvironment depends on the intricate competition between nutrient uptake, cooperated by its flow rate, and abiotic stress: here, flow induced hydrodynamic stress,

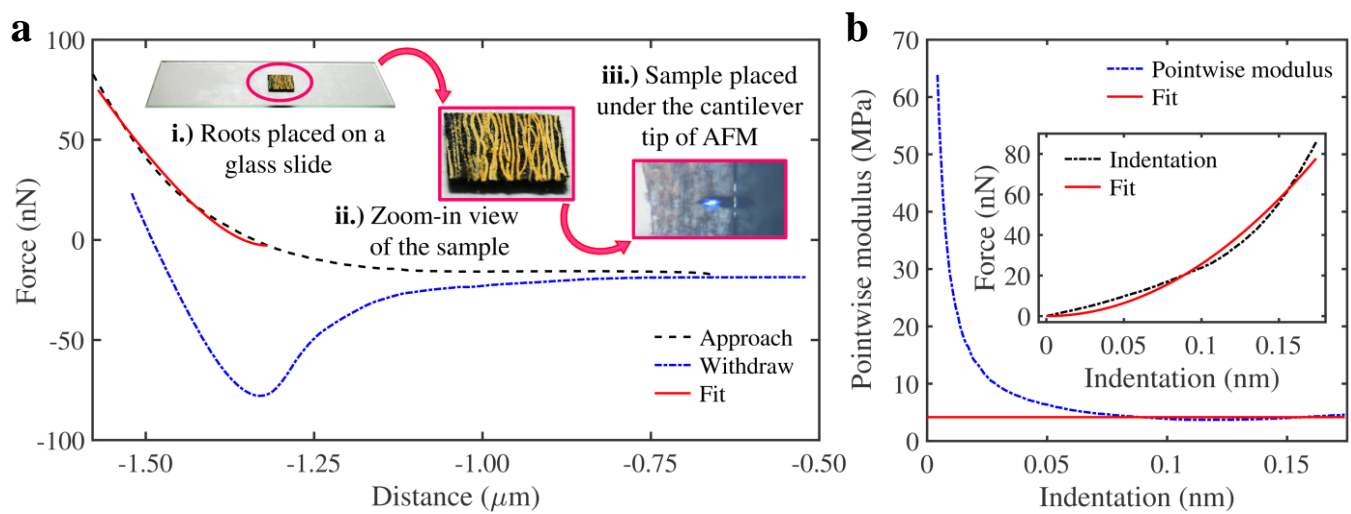

**Fig. 4** (a) Force-distance curve obtained from AFM analysis and then fitted to Hertz model to extract the Young's modulus of the root sample. (A: i-iii) Roots of *Brassica juncea* in post-germination stage, were cut into pieces (1cm), dried overnight and then placed on a microscopic glass slide to minimize spacing between them. The air-dried root sample was imaged using an Asylum Research MFP-3D-BIO (Asylum Research of Oxford Instruments) Atomic Force Microscope (AFM). AFM imaging was conducted in Nanoindentation mode using a silicon cantilever probe with a radius of 7 nm and a spring constant 26 k (N/m), (Model: AC160TS-R3, Oxford Instruments). (b) and inset within represent the point-wise modulus and indentation curves obtained from the analysis of AFM data using AtomicJ software. The Young's modulus values obtained were subsequently utilized in simulations conducted in COMSOL Multiphysics Software™.

resulting to thigmomorphogenesis. Understanding this complex interplay between nutrient flow and root structure interaction, which is typical in the growth of plant roots in nutrient-mediated flow environments, is of particular importance. To delve deeper into this competition, we developed a mathematical modeling framework and performed numerical simulations on nutrient flow-root interaction using real-time images of the root captured from our experiments at a given temporal instant. Young's Modulus, a critical parameter for simulations of the growing root inside the channel, was determined by curve fitting (Fig. 4a and 4b) obtained from

Atomic Force Microscopy (AFM), as elaborately discussed in the "*Materials and methods*" section.

The simulated results illustrate contours representing effective internal stress, normal loading at the interface, and the magnitude of loading, all of which indicate average hydrodynamic stress levels (Fig. 5a). These simulated results are associated with the anatomical sections depicted in the last column of Fig. 5a, which reveal minimal differences in both the structure and area (A($Q_1$) = 0.2299±0.0025[a] mm$^2$, A($Q_2$) = 0.2311±0.0021[a] mm$^2$, A($Q_3$) = 0.2287±0.0026[a] mm$^2$) of the vascular bundles within the root. A schematic representation of the anatomical section is shown in Fig. 5b, depicting the cellular architecture. The mean values, A($Q_1$), A($Q_2$) and A($Q_3$), do not differ significantly at $p_r \leq 0.01$ according to Duncan's multiple range test (Fig. 5c).

The average stress (inside the root, $\sigma_{avg}$) experienced by the growing root in the PRFD varies from $8.05 \times 10^{-3}$ N/m² to $166.42 \times 10^{-3}$ N/m² for the flow rates 0.05 mL/hr to 1.2 mL/hr (Fig. 6a). The gradual increase in flow rates induces greater fluid flow-induced stress on the soft tissues of growing root. In Regime-I of Fig. 6a, which maps the flow rates from $Q$ = 0.05 mL/hr to 0.04 mL/hr, the average stress exhibited a steeper slope of 0.1659 (N/m²)/(mL/hr). In Regime-II of Fig. 6a, mapping flow rates beyond 0.4 mL/hr, specifically, $Q$ = 0.6 mL/hr to 1.2 mL/hr, the average stress exhibited a relatively lesser steep slope of 0.1288 (N/m²)/(mL/hr) than that of Regime-I. These variations in stress are reflected in the anatomical changes of the root, i.e.,

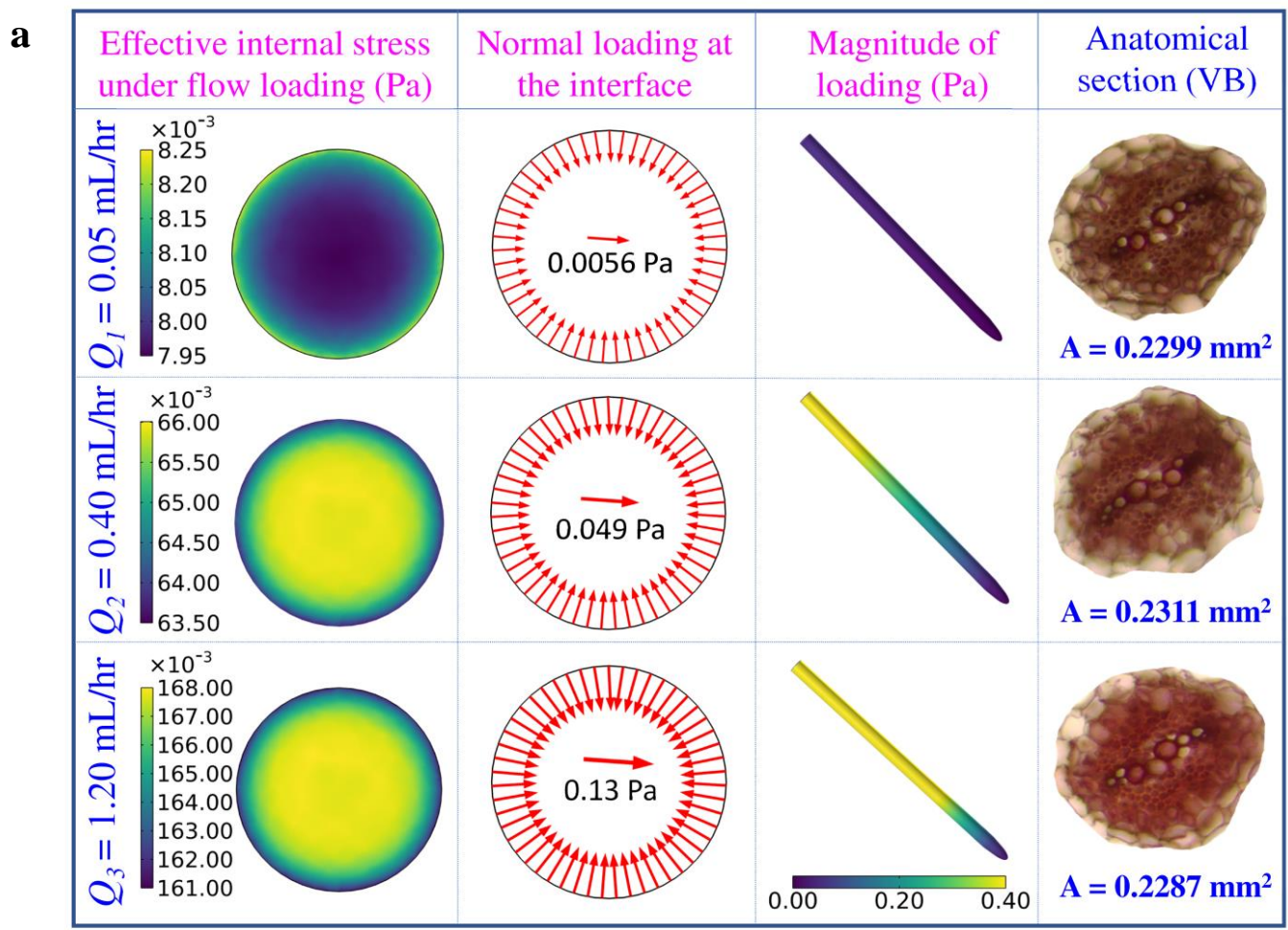

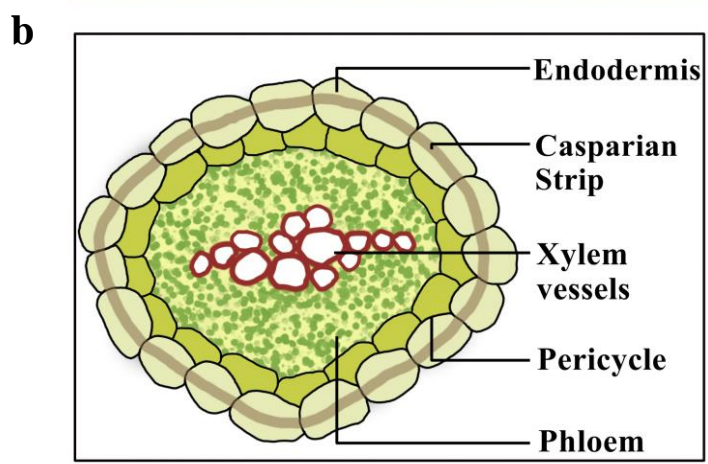

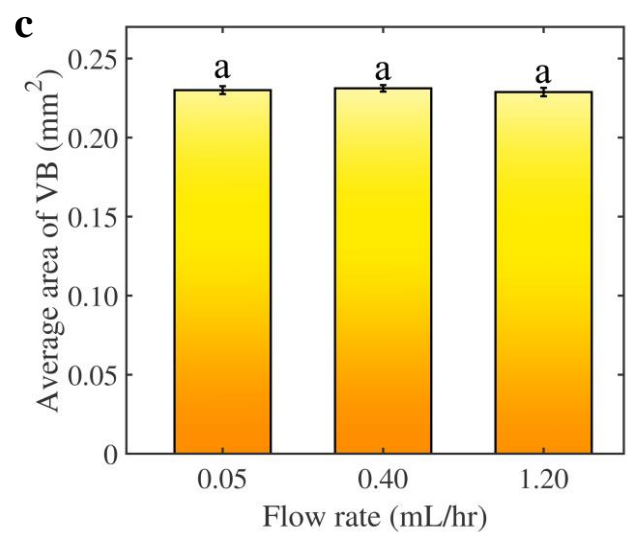

Fig. 5 (a) Contours of effective internal stress under flow loading (Pa), normal loading at the interface, and magnitude of loading (Pa) obtained from numerical simulations; Anatomical sections (extreme right) represent the area of vascular bundles (VB) with no significant structural and area difference. (b) Schematic representation of the cellular architecture shown in Fig.4a (c) Variations in the average area of vascular bundle at different flow rates $Q$ = 0.05 mL/hr, $Q$ = 0.4 mL/hr and $Q$ = 1.2 mL/hr. The graphs were plotted using MATLAB® R2022a and the significance of the data was verified performing one-way ANOVA using IBM® SPSS® software. The mean values marked with the same letter(s) do not differ significantly at p_r≤0.05 according to Duncan's multiple range test.

percentage reduction (R) of the average area of cortical cells for flow rates $Q_2$ and $Q_3$ relative to the lower flow rate $Q_1$, as shown in the inset of Fig. 6a. We observed a 4.159 % reduction ($R_{Q_2,Q_1}$) in the area between cortical cells of roots subjected to flow rates $Q_2$ and a 21.580 % reduction ($R_{Q_3,Q_1}$) in area for roots subjected to flow rates $Q3$. These changes, along with the change in the number of cortical cells (Fig. 3a) indicate thigmomorphogenesis pertaining to the hydrodynamic stress experienced by the growing root with an increase in flow rates from $Q_1$ to $Q_3$. It is to be noted here that though changes in cortical cell count and area were observed with the change in flow rates (Fig. 3a), no significant change in the structure and area of vascular bundles within the roots, critical for water uptake and transport, was observed (Fig. 5a-last column). These observations indicate the shielding effect provided by cortical cells to the inner vascular bundles against flow-induced stress.

**Flow Modulated Stress versus Nutrient Uptake**

The analysis of macronutrient, nitrogen, was performed using the TKN method. The overall data from the Kjeldahl analysis revealed a trend where nitrogen uptake increased with an increase in the flow rate. In Regime-I of Fig. 6b, which covers flow rates from $Q$ = 0.05 mL/hr to 0.04 mL/hr, including no-flow condition, the nitrogen percentage exhibited a steep slope of 2.765 % $N_2$/(mL/hr). This steep slope signifies a gradual and linear increase in nutrient uptake induced by the flow. Interestingly, this linear increase continues until the optimum flow rate, $Q$ = 0.4 mL/hr, beyond which the slope becomes less steep. In Regime-II of Fig. 6b, encompassing flow rates beyond 0.4 mL/hr, specifically $Q$ = 0.6 mL/hr to 1.2 mL/hr, the nitrogen percentage exhibited a relatively lesser steep slope of 0.665 % $N_2$/(mL/hr). The lesser steepness of the slope in Regime-II of Fig. 6b reflects the overpowering effect of the stress induced by the further increase in flow rate beyond $Q$ = 0.4 mL/hr (cf. Fig. 6a). This indicates that while nutrient uptake initially increases with flow rate, there is a threshold beyond which the stress induced by higher flow rates begins to negatively impact nutrient uptake, leading to a less steep increase in nitrogen percentage. It is indeed fascinating to observe that the stress induced by the increase in flow rate is mitigated by the cortical cells, thereby leaving the vascular bundle region almost unaffected (Fig. 5a: witnesses that the vascular bundle area, shown in the last column, remains almost unchanged with increasing flow rate). However, during this process, the cortical cells experience a reduction in average cell area (Fig. 3c and inset of Fig. 6a). This reduction in cell area is compensated for by an increase in the number of cortical cells (Figs. 3A and B, witness that $C_c$ increases with flow rate), contributing to the overall increase in nitrogen uptake in Regime-II, though at a lesser rate compared to that in Regime-I of Fig. 6b.

In 1973, Jaffe [33] conducted a study involving a variety of plant species, wherein he applied gentle mechanical stimulation by rubbing the internodes and observed a noteworthy reduction in stem elongation as a direct response to the mechanical stimulus. This investigation sheds light on the suppressive impact of mechanical stimulation, particularly through rubbing, on stem elongation within specific plant species. This phenomenon is characterized as "thigmomorphogenesis" and is perceived as an adaptive mechanism aimed at mitigating environmental stresses, with potential involvement of the hormone ethylene. Thus, thigmomorphogenesis exerts a significant influence on various facets of plant growth and development. It is worth emphasizing that the character and extent of this response hinge on factors such as the plant's species or variety and the specific physiological stage at which the plant undergoes stimulation. Our study advances the understanding of root-cell behavior in *Brassica juncea* plant using a microfluidic platform, with a focus on the cooperative correlative influence of fluid-flow-assisted nutrient uptake and corresponding mechanical pressure as stimuli, resulting in what we term 'flow-induced thigmomorphogenesis'.

The water and nutrients absorbed by the root epidermis are transported by the cortex, located between the root epidermis and endodermis, into the vascular bundle by apoplastic and symplastic routes, and are involved in determining the fate of epidermal cells [62]. The cortex in plants is a relatively undifferentiated cell type called ground tissue. Though the cortex layers vary among species, they remain almost unchanged at a particular developmental stage for a specific species [63]. However, under stressed conditions, such as mechanical stress [64], cortical cell division and root developmental plasticity [65] are regulated by auxin, a phytohormone that primarily controls growth and development patterns in plant [66]. Auxin is known to accumulate at the site of stress through long- and short-distance auxin flow [65], resulting in cortical cell division. Our experimental findings and simulated data corroborate the enhancement of stress with an increase in flow rates (cf. Fig. 6a). Quite remarkably, as evident from Fig. 3a, an increase in flow rate ($Q_1$ to $Q_3$) results in a significant increase in the number of cortical cells, primarily attributed to the stress-stimulated auxin accumulation in the roots as referred to earlier.

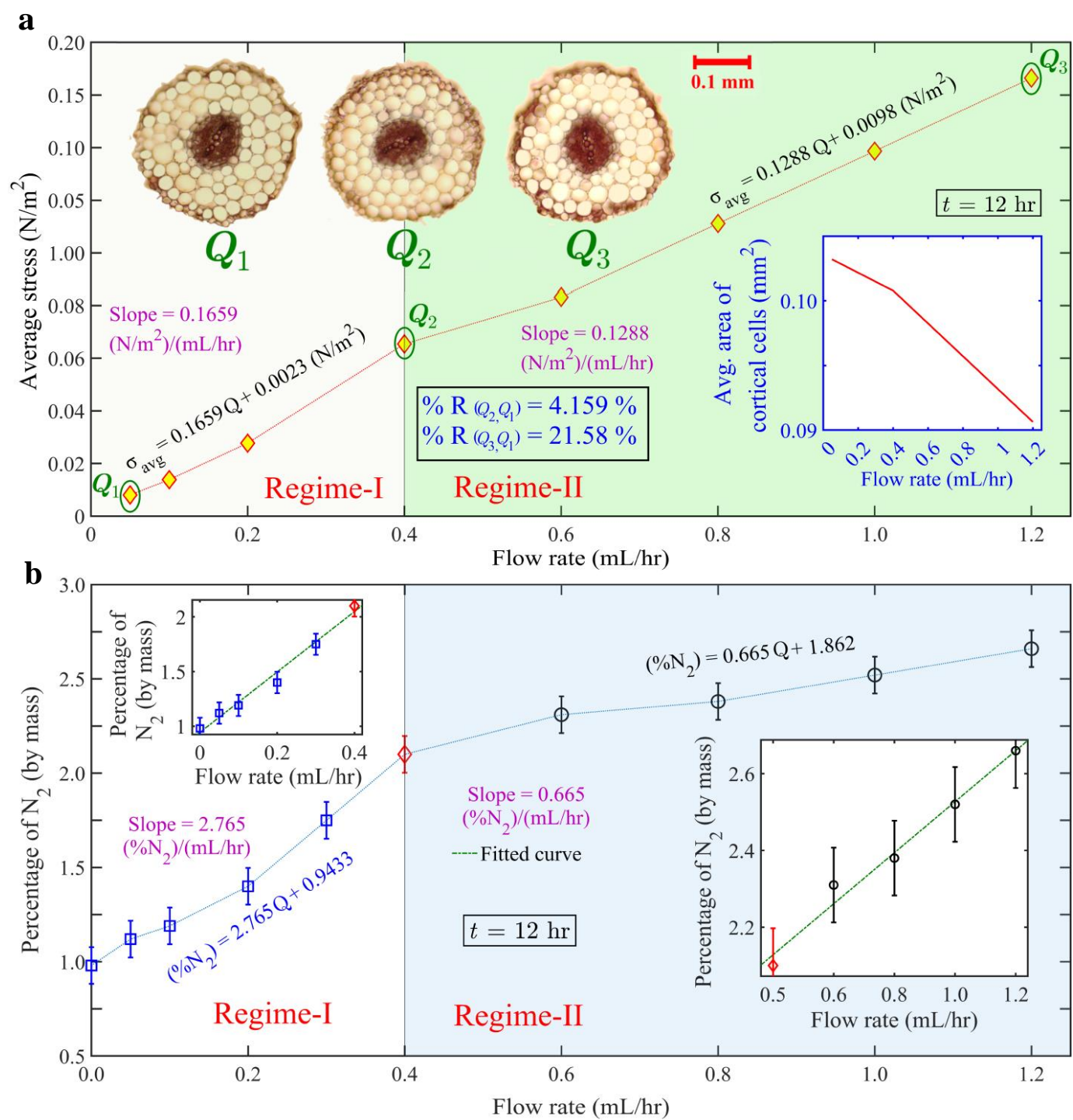

**Fig. 6** (a) Variation in average stress (N/m$^2$), obtained from numerical simulations, exerted by the nutrient solution on the root growing inside the microfluidic channel with respect to different flow rates ($Q$) at t = 12 hr. $Q_1$, $Q_2$ and $Q_3$ denote the flow rate: 0.05 mL/hr, 0.4 mL/hr and 1.2 mL/hr, respectively. The anatomical sections for roots subjected to flow rates $Q_1$, $Q_2$ and $Q_3$ are provided in the inset. The percentage reduction (R) in the area of individual cortical cells is calculated with respect to the flow rate, $Q_1$, for $Q_2$ and $Q_3$. A 4.159 % reduction in the area is observed between cortical cells of the root subjected to flow rates $Q_1$ and $Q_2$, whereas a 21.580 % reduction in area is observed for root subjected to flow rates $Q_1$ and $Q_3$, the change in trend of the area of cortical cells is represented in inset. The two regimes, namely, Regime-I and Regime-II, are based on the slope of the average stress with flow rate ($Q$). The equations for the slope of the corresponding flow rates obtained from the best curve fitting are obtained as: average stress inside the root ($\sigma_{avg}$) = 0.1659 $Q$ +0.0023 in N/m$^2$ and $\sigma_{avg}$ = 0.1288 $Q$ + 0.0098 in N/m$^2$, with $Q$ in mL/hr, for Regime-I and Regime-II, respectively. (b) Variations in the percentage of N$_2$ extracted from root samples with an equal mass (0.2 g) in relation to changes in flow rate. Total nitrogen percentage was determined using the Total Kjeldahl Nitrogen (TKN) method. The two regimes, namely, Regime-I and Regime-II, are based on the slope of the percentage of nitrogen with flow rate ($Q$). The slope of the corresponding flow rates obtained from the best curve fitting is depicted in the inset, and the corresponding equations are obtained as: % N$_2$=2.765 $Q$ +0.9433 and % N$_2$=0.665 $Q$ +1.862, with $Q$ in mL/hr, for Regime-I and Regime-II, respectively.

In our research, we have, for the first time, investigated the impact of mechanical stimuli induced by variations in flow rates on the morphological and anatomical characteristics of *Brassica juncea* plant roots. Our findings align with the well-established concept of thigmomorphogenesis, a term frequently employed to elucidate plant responses to mechanical stimuli. The changes observed in our study, which encompass modifications in root structure and the outcomes of stress-related fluid-root interaction simulations, underscore the dynamic and adaptive nature of thigmomorphogenesis in plants undergoing alterations in mechanical forces.

## Conclusion

This study introduces a novel plant root fluidic device designed for studying flow-induced thigmomorphogenesis in germinating roots. The developed fluidic device, whose hydraulic conductivity conforms to that of the soil, serves as an inert, easy-to-use, and cost-effective platform to analyze individual root samples and enables seamless real-time monitoring of root morphology. Findings reveal that increasing flow rates boost root length and nitrogen uptake until reaching an optimum flow rate, beyond which flow induced stress reduces root length as aptly confirmed by the results of this endeavor. However, roots in flow conditions consistently outperform those in no-flow conditions due to enhanced nitrogen uptake. The study concludes that nutrient flow induces morphological changes, offering insights into root adaptation and cellular responses. The novelty of this research lies in its exploration of flow-induced thigmomorphogenesis in early root growth, the development of a novel microfluidic system, the investigation of the nitrogen-root growth relationship, the integration of simulation and quantitative analysis, and the high-resolution insights it provides. These contributions collectively advance our understanding of underlying nutrient flow-plant root interactions and have practical implications for agriculture and environmental science. Future studies could explore the molecular mechanisms underlying phenotypic changes related to different flow rates. A deeper understanding of the cellular and molecular processes involved in flow-induced thigmomorphogenesis could inform the design of resilient hydroponic systems and support soil-less crop production.

## Author Contributions

K.A., Conceptualization, data curation, visualization, formal analysis methodology, investigation, experimentation, writing - original draft; S.K.M., Conceptualization, formal analysis, visualization, data curation, methodology, writing - original draft, software; P.K.M., Conceptualization, methodology, supervision, writing – review & editing, project administration, funding acquisition, resources.

## Conflicts of interest

There are no conflicts to declare.

## Acknowledgements

This work is supported by the Science and Engineering Research Board (SERB), Govt. of India, under the project no. CRG/2022/000762. The Ph.D. fellowship of Kaushal Agarwal is supported by the Ministry of Education (MoE), Govt. of India. The authors thank the TIDF (TIH) Indian Institute of Technology, Guwahati for their active support. The authors are grateful to CIF, IIT Guwahati for the instrumentation facility. The authors greatly acknowledge Dr. Sudha Gupta, Department of Botany, University of Kalyani, India for her help with the discussion pertaining to the plant biology.